\documentclass[11pt,oneside,a4paper]{article}

\usepackage{natbib}
\usepackage{chngcntr}
\usepackage{lineno,hyperref}
\usepackage{rotating}
\usepackage{amsfonts}% per usufruire di mathbb{}%
\usepackage{amsmath}
\usepackage{amsthm}
\usepackage{xcolor}
\usepackage{amssymb,amsmath}
\usepackage{subfig}
\usepackage[english]{babel}
\usepackage{graphicx}
\usepackage{tabularx}
\usepackage{booktabs}
\usepackage{enumerate, hyperref}
\usepackage{epstopdf} % EPS to pdf conversion

\usepackage{tikz,tikz-inet,pgf,pgfplots}
\pgfplotsset{compat=1.5}

\newcommand{\figuretextsize}{\footnotesize}

\theoremstyle{plain}

\theoremstyle{definition}

\theoremstyle{remark}

\title{``Chaos'' in energy and commodity markets: a controversial matter}

\author{Loretta Mastroeni$^1$
   \and Pierluigi Vellucci$^1$}

\begin{document}

\maketitle

\begin{abstract}
We test whether the futures prices of some commodity and energy markets are determined by stochastic rules or exhibit nonlinear deterministic endogenous fluctuations.
As for the methodologies, we use the maximal Lyapunov exponents (MLE) and a determinism test, both based on the reconstruction of the phase space. In particular, employing a recent methodology, we estimate a coefficient $\kappa$ that describes the determinism rate of the analyzed time series. We find that the underlying system for futures prices shows a reliability level $\kappa$ near to $1$ while the MLE is positive for all  commodity futures series. Thus, the empirical evidence suggests that commodity and energy futures prices are the measured footprint of a nonlinear deterministic, rather than a stochastic, system.

\end{abstract}

\section{Introduction}
\label{sec:intro}

Over the years, chaos{\let\thefootnote\relax\footnotetext{Keywords: chaos, butterfly effect, commodity futures.}} theory{\let\thefootnote\relax\footnotetext{$^1$Dept. of Economics, Roma Tre University, via Silvio D'Amico 77, 00145 Rome, Italy; loretta.mastroeni@uniroma3.it, pierluigi.vellucci@uniroma3.it.}} has gradually provided a framework to study some interesting properties of time series. An important reason to be interested in chaotic behaviour is that it can potentially identify, among many time series, which of them appear to be actually random. In particular, as for commodity as well as many other markets, evidence on deterministic chaos would have important implications for regulators and short-term trading strategies.

Moreover, if the random walk model is not a proper account of public  market behaviour, then the debated  Efficient Market Hypothesis, that plays such a basic role in the  markets, might be meaningless in this context.

The question is whether such random-looking data are actually random or completely deterministic. If they are completely random, then their behaviour is not predictable anyway; otherwise, it is possible to predict their behaviour over short periods of time (whereas long prediction is impossible, due to the instability of chaotic systems). Hence, this distinction provides the predictability degree of the analyzed system.

The presence of chaos in the time series of commodity and energy markets is a controversial matter: one can take a look at the literature on energy markets, where the evaluations performed over the years show contrasting results, which might be justified, for example, by a misunderstood application of the theoretical aspects of chaos theory.

Chaos theory is linked to \cite{Devaney89}, but in general the literature actually deals with the  \emph{butterfly effect}, which, according to Devaney, is only one of the properties of the definition of chaos.

The butterfly effect entails that, if two initial conditions slightly differ for a quantity $\delta x$, their difference after time $\delta t$ will be $\delta x e^{\lambda \delta t}$ with $\lambda > 0$, that is, with exponential separation. Small differences in initial conditions (such as those due to rounding errors in numerical computation) yield widely diverging outcomes for such dynamical systems, making long-term prediction impossible in general.

%The aim of this paper is twofold. First, we perform a brief literature review of commodities markets enlightening the role of sensitive dependence on initial conditions in chaos definition: the mathematical definition of chaos recalled here (according to Devaney \cite{Devaney89}) is helpful to prevent us from misleading results about ostensible chaoticity of the returns series. %Second, we test for sensitive dependence on initial conditions introducing a coefficient $\kappa$ that describes the \emph{determinism rate} of the analyzed time series, which represents, in percentage, its reliability level. The introduction of this reliability level is motivated by the fact that, as we have already said, time series generated from stochastic systems might show sensitive dependence on initial conditions.

Butterfly effect can be checked by an entropy test (\cite{farmer1982chaotic}, \cite{grassberger1983estimation}) which employs the Kolmogorov entropy (\cite{kolmogorov1958new}) and the maximal Lyapunov exponents (MLE) (\cite{Schu89}, \cite{Ott93}, \cite{Stro94}, \cite{Ko05}).

%The results obtained with the maximum Lyapunov exponent should indicate the presence of butterfly effect. Nevertheless, this phenomenon --- which is usually showed by deterministic systems --- here is not found to be completely deterministic. This result is in accordance with some experiments documented in physics; in fact, it is argued that some time series generated from stochastic systems may show sensitive dependence on initial conditions (see \cite{tanaka1996lyapunov}, \cite{ikeguchi1997lyapunov}, \cite{tanaka1998analysis}).

%The presence of butterfly effect in commodity futures markets is a controversial matter, but the evaluations obtained here ...  In other words, commodities considered here encompass both a stochastic contribute and a deterministic one, which .... (see Section \ref{sec:exp}).
%%%%%%%%%%%%%%%%%%%%%%%%%%%%%%%%%%%%%%%%%%%%%%%%%%%%%%%%%%%%%%%%%%%%%%%%%%%%%%%%%%%%%%%

We analyze the nonlinear deterministic structure in some commodity and energy markets by testing for sensitive dependence on initial conditions. Our data set consists of daily prices of commodities (natural gas, heating oil, gold, silver, corn, oats, cocoa, coffee, feeder cattle, lean hogs), considered in two previous papers \cite{benedetto2015maximum, benedetto2016predictability} covering several ranges in the period 01.07.1959 - 15.05.2014.

As for the methodologies, we use the Lyapunov Exponents and a determinism test, both based on the reconstruction of the phase space. In particular, we employ a coefficient $\kappa$ that describes the \emph{determinism rate} of the analyzed time series. The coefficient $\kappa$ represents, in percentage, the reliability level about the test on the sensitive dependence on initial conditions. The introduction of this reliability level is motivated by the fact that time series generated from stochastic systems might show sensitive dependence on initial conditions (see \cite{tanaka1996lyapunov}, \cite{ikeguchi1997lyapunov}, \cite{tanaka1998analysis}). The coefficient $\kappa$ has been introduced by \cite{Ka92}, and here it is estimated  by employing a recent methodology developed by \cite{Ko05}.

In our work, the reliability level $\kappa$ yields results near to $1$ while the MLE is positive for all  commodity futures series. This means that they show a considerable contribution of determinism. In this way, we can ensure the presence of butterfly effect (as specified above, one of the properties of a chaotic system,
according to Devaney \cite{Devaney89}) in the commodity futures markets considered in the paper.

This study contributes to an overall picture of the role of chaos in the energy and commodity markets. In particular, we define a working hypothesis that addresses three important features of chaotic signals, namely, the existence of a low-dimensional attractor in the underlying dynamics, the presence of sensitive dependence on initial conditions, and the deterministic behaviour of the system. The methodologies we use suggest that commodity futures prices are the measured footprint of a deterministic, rather than stochastic, system. Determinism analysis suggests that there are several deterministic forces interacting with each other. The presence of a chaotic dynamics could be connected to the existence of several deterministic forces that may result in complex price movements in financial markets. Today, the complexity of financial and commodities markets is very high because world decisions in business, finance and economics are influenced by sociologic, environmental, and geopolitical factors. In this regard, Panas and Ninni \cite{Panas00} write in their still relevant conclusions: ``An energy economist who is interested in the dynamic behaviour of the complex system that governs the oil markets needs to know how sensitive the system is to initial conditions, and to achieve this he needs to estimate the Lyapunov exponents''.

Determinism is related to the role of information in the markets that, no doubts, is of paramount importance. Let us think to the stepping formalization proposed by Fama (see e.g. \cite{malkiel1970efficient}, \cite{Fama1998283}). The central issue is whether or not to adopt trading strategies that achieve excess returns relative to the market, based on information contained in the historical data. Currently, the empirical evidence would seem to indicate that markets are often not efficient, even in weak form. The perception of a trend as seemingly stochastic could be due to the lack of knowledge of the information underlying it.

%%%%%%%%%%%%%%%%%%%%%%%%%%%%%%%%%%%%%%%%%%%%%%%%%%%%%%%%%%%%%%%%%%%%%%%%%%%%%%%%%%%%%%%
The plan of the paper is as follows. Section \ref{sec:review} provides a brief review of chaos theory results in the field of commodity futures markets. In Section \ref{sec:intro2} we consider the implications of chaos in commodity futures markets. In Section \ref{sec:exp} we describe the dataset and present estimate of three diagnostic tests for deterministic butterfly effect: (i) reconstruction of the phase space, where we estimate the smallest sufficient embedding dimension of the system using the FNN algorithm; (ii) Lyapunov exponents, which measure the divergence rate; (iii) determinism test, to distinguish deterministic behaviour from stochastic one. In Section \ref{sec:disc} we discuss and interpret the estimates obtained in Section \ref{sec:exp}, comparing them with the results found in the literature. Finally, Section \ref{sec:conc} is devoted to the conclusions.

\section{Brief literature review}
\label{sec:review}

%During the last 20 years, researchers have shown their great interest throughout the new ways of using elements of the so-called ``chaos theory'' to analyze economic and financial time series. According to this theory, very complex behaviours of financial series, which appear to be random, may be explained by a deterministic nonlinear system. In particular, we recall the following results in the field of commodity futures markets.

The presence of butterfly effect in commodity futures markets is a controversial
matter, as can be deduced from the literature reviewed below: some papers claim they have detected the presence of chaos (butterfly effect), while others state the opposite.

\cite{chwee1998chaos} tests for the presence of butterfly effect using the NYMEX 1-month, 2-month, 3-month, and 6-month daily natural gas settlement prices, from April 1990 to September 1996. The results fail to provide significant evidence of butterfly effect. \cite{serletis1999north} test for butterfly effect in seven Mont Belview, Texas hydrocarbon markets, using monthly data from 1985:1 to 1996:12 (for the markets of ethane, propane, normal butane, iso-butane, naptha, crude oil, and natural gas) and find an evidence of butterfly effect.

\cite{Panas00}investigate butterfly effect in daily price data for two major petroleum markets, namely those of Rotterdam and of  Mediterranean. The sample consists of the daily prices of different oil products from 4 January 1994 to 7 August 1998, resulting in 1161 observations. All prices were collected from OPEC. The main results obtained by the authors are summarised in their Table 5. They claim to show strong evidence of butterfly effect in a number of oil products considered.

\cite{Adrangi01} investigate the presence of butterfly effect in crude oil, heating oil, and unleaded gasoline futures prices from the early 1980s. Daily returns data from the nearby contracts are diagnosed by employing correlation dimension test, the BDS test and Kolmogrov entropy. They find strong evidence of non-linear dependence in the data, but it is not consistent with chaos.

The study performed by \cite{Panas01} analyzes the daily pricing for nonferrous metals (aluminium, copper, lead, tin, nickel and zinc), considering daily closing metal prices over the period from January 1989 to December 2000 (the number of observations is 2987 and the tin series begins in August, 1989) and finding evidence of ``deterministic chaos only in the case of tin returns''.

\cite{adrangi2002dynamics} employ daily prices of the nearby (expiring) palladium and platinum futures contracts traded on The Commodity Exchange from November 1983 through March 1995 and January 1975 through June 1995, respectively, focusing their tests on daily returns. They find that the nonlinearity in palladium and platinum is inconsistent with chaotic behaviour.

\cite{Chatrath02} conduct tests for the presence of low-dimensional chaotic structure in the futures prices of four agricultural commodities: soybean (from 1969 to 1995), corn (from 1969 to 1995), wheat (from 1968 to 1995), cotton (from 1972 to 1995). Even though there is strong evidence of non-linear dependence, this suggests that there is no long-lasting chaotic structure.
\cite{moshiri2006forecasting} examine daily crude oil futures prices from 1983 to 2003, listed in NYMEX; their test provides negative evidence of butterfly effect. \cite{matilla2007nonlinear} studies the butterfly effect nature of three energy futures series --- natural gas, unleaded gasoline and light crude oil --- finding evidence in futures returns.

\cite{sakai2007transition} investigate piglet-pricing data in Japan, considering the monthly data for the real price and population of piglets over 1967 and 1992. An application of Lyapunov spectrum analysis to the data is carried out in order to distinguish deterministic chaos and periodic solutions. Their analysis shows that government intervention might reduce market instability by removing a chaotic market's long-term unpredictability.

\cite{kyrtsou2009energy}, analyze five energy products (crude oil, gasoline, heating oil, propane, and natural gas) over the period from 1994 to mid-January 2008. They reject the null hypothesis of butterfly effect behaviour.

\cite{barkoulas2012metric} consider a data set which consists of daily oil spot prices covering the period 1.2.1985 - 8.31.2011. They do not find butterfly effect tendencies in the oil market, suggesting that ``oil spot prices are the measured footprint of a stochastic rather than a deterministic system'' (pp. 585).

\section{Concepts and implications of butterfly effect in commodity markets}
\label{sec:intro2}

Several studies show evidence of nonlinearity for various financial time series: \cite{barnett2000martingales}, \cite{franses2000non}, \cite{sarantis2001nonlinearities} and \cite{zhang2001nonlinear}, among others.

Nonlinear dynamics may be able to explain a large set of time series behaviours. One motivation for this line of research is to determine whether the non-linearities are consistent with chaotic time paths, which have several properties that would be of special interest for commodity market observers (for instance, an apparent stochasticity of time series that could be generated by deterministic systems).

In the literature the interest in whether a prices series is chaotic has been focused on the debate, inter alia, on the worth of the forecasting/technical analysis in the very short run. In fact, a couple of decades ago, several studies showed that this (nonlinear) analysis may provide better results in predicting the price behaviour of many financial instruments. See for instance, among others: \cite{osler1995head}, \cite{clyde1997charting} and references therein. These works are a little dated and deal with financial instruments (not commodities), but are also very interesting since they link the success of technical analysis in the short time and the chaoticness of time series.

\cite{osler1995head} examine a technical strategy that can be viewed as one of a large class of nonlinear prediction rules potentially deriving from nonlinear versions of structural models such as the monetary models, chaos models, and many others. ``Many of these models have been shown to fit the data with some acceptable level of explanatory power within sample, and some appear to be helpful in forecasting conditional exchange rate variances. Nonetheless, out of sample tests of these models indicate that they generally forecast short-term exchange rate changes with little or no greater success than the random-walk model'' ( \cite{osler1995head}). \cite{clyde1997charting} simulate a ``chaotic'' (we would say ``sensitive to initial conditions'') price series and prove that in Head-and-Shoulders trading strategies there is considerable evidence of the fact that technical analysis does work better on nonlinear data than on random data, but the frequency of ``hits'' (successes) employing the technical trading rule become comparable after just a few trading days (\cite{clyde1997charting} Table IV). Thus, there is evidence that short-term trading could benefit from knowing whether or not the price series are affected by butterfly effect.

\begin{figure}[tb]
\centering
\includegraphics[scale=0.60]{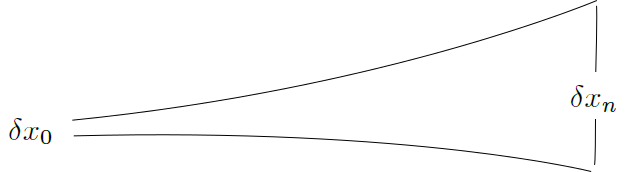}
\caption{Effect of a small change of initial condition.}
\label{fig:1}
\end{figure}

Let us try to give a basic motivation in order to explain, from a purely theoretical point of view, the above empirical reasoning.
Let us consider the time series $y_n$ and assume that there exists a system $(g,f, x_0)$ such that $y_n=g(x_n)$, $x_{n+1}=f(x_n)$, where $x_0$ is the initial condition at the initial time $n=0$, $g$ maps the $m$-dimensional phase space $\mathbb R^m$ to $\mathbb R$, and $f$ maps $\mathbb R^m$ to $\mathbb R^m$. The function $f$ maps an unknown (to the econometrician) dynamics that governs the evolution of the unknown (to the econometrician) state $x_0$. The econometrician observes $y_n$. The task is to uncover information about $(g,f, x_0)$ from observations $y_n$. %\cite{brock1988business}.
The time series $y_n$, which we will assume as the data time series under analysis, has a chaotic explanation if $x_n$ is chaotic.
The question is whether it is possible to forecast a chaotic series. Intuitively, the butterfly effect usually does not allow long-term forecasting of chaotic series. If we change slightly the value of the initial point: $x_0$ $\mapsto$ $x'_0=x_0+\delta x_0$, the point $x_n$ at discrete time $n$ will also be changed (see Fig. \ref{fig:1}).
What may happen is that, when time becomes large, the small initial distance $\delta x_0$ grows anyway, and it may grow exponentially fast: $\delta x_n\sim \delta x_0 e^{\lambda n}$, for some $\lambda>0$. The term $\delta x_n$ represents the uncertainty induced by perturbations. Hence, fixed $\delta x_0$ and $\lambda$, it is bounded if $e^{\lambda n}$ is bounded and so if $n$ is as small as possible ($n=1$ or $n=2$, for instance): this is the short-term forecasting of chaotic series, which is possible because $\delta x_n$ is amplified at a finite rate $e^{\lambda n}$. Accordingly, for chaotic time series, if one knows $(g,f)$ and could measure $x_n$ without error, one could perfectly forecast $x_{n+i}$ and, thus, $y_{n+i}$ in the short time (let's say a few days when dealing with daily data).

\section{Methodology and dataset}
\label{sec:exp}

\begin{table}
\centering\begin{tabular}{lll}
\toprule
Commodity contract & Temporal range & Number of points \\
\midrule
Natural gas  &03.04.1990-15.05.2014 & 6.042\\
Heating oil &06.03.1979 - 15.05.2014 & 8.825\\
Gold & 31.12.1974 - 14.05.2014 & 9.884\\
Silver & 13.06.1963 - 15.05.2014 & 12.758\\
Corn & 01.07.1959 -15.05.2014 & 13.817\\
Oats & 01.07.1959 - 15.05.2014 & 13.817\\
Cocoa &05.01.1970 - 15.05.2014 & 11.097\\
Coffee& 17.08.1973 -15.05.2014 & 10.194 \\
Feeder cattle &06.09.1973- 15.05.2014 & 10.258\\
Lean hogs &25.06.1969 - 15.05.2014 & 11.297\\
\bottomrule
\end{tabular}
\caption{Description of the data set.}
\label{table:dataset}
\end{table}

In this paper we test for sensitive dependence on initial conditions and determinism of the following futures series: two energy series (natural gas, heating oil), two metal series (gold, silver), two grains (corn, oats), two soft commodities (cocoa, coffee) and two other agricultural commodities (feeder cattle, lean hogs) futures from the Chicago Board of Trade (CBOT), Chicago Mercantile Exchange (CME), Inter Continental Exchange (ICE), New York Mercantile Exchange (NYMEX), and its division Commodity Exchange (COMEX). The time series were obtained from \url{http://www.quandl.com}. Further information on the data set is provided in the Table \ref{table:dataset}. The series were already considered in \cite{benedetto2015maximum, benedetto2016predictability}, where the predictability of commodity market time series is investigated by predicting their entropy (for a surveys on the topic see \cite{benedetto2015signal}).

By way of the following subsections, we explain step by step the methodology of the paper.

\subsection{State space reconstruction and the embedding dimension ($m$)}

Let the price of commodity at time $t$ be denoted by $p_t$, then returns are measured as $z_t=\ln\frac{p_t}{p_{t-1}}$. A scalar time series $\{z_t, t=1,2,\dots,n\}$, $n\in\mathbb N$ represents the observations from the markets examined in this paper, which can be used to reconstruct the state space (the so-called ``phase space'') \cite{Pack80}. The phase space is defined as the multidimensional space whose axes consist of variables of a dynamical system. The asymptotic behaviour of the dynamical system is related to an \emph{attractor}, whose dimension will provide a measure of the minimum number of independent variables able to describe the dynamical system.
The state space reconstruction is the fundamental step for recovering the properties of the original attractor from a scalar time series. In this respect, it is possible to prove that an attractor, which is topologically equivalent to the scalar time series, can be reconstructed from a dynamical system of $N$ variables by using a method of time delay coordinate (\cite{Take81}, \cite{ruelle1989chaotic}). The reconstructed attractor of the original system is given by the vector sequence
\begin{equation}
\label{eq:3}
\textbf{p}(i)=\left(z_i,z_{i+\tau},z_{i+2\tau},\dots,z_{i+(m-1)\tau}\right)
\end{equation}
where $\tau$ is an appropriate time delay, $m$ is the embedding dimension and index $i$ varying on $\{1,2,\dots, n-(m-1)\tau\}$.

Estimation of the smallest sufficient embedding dimension has been performed through the method of the \emph{false nearest-neighbours algorithm} (\cite{Kennel92}). Here, the minimum embedding dimension is such that any further increase in the dimension does not significantly increase the distance between any two neighbouring points in the trajectory. If we use the maximum norm, the statistics providing the fraction of false neighbors is \cite{kantz2004nonlinear}
\begin{equation}
    \textrm{FNN}(r) = \frac{\sum_{i=1}^{n-m-1}\Theta \left( \frac{\vert \underbar{s}_{i}^{(m+1)}-s_{k(i)}^{(m+1)}\vert}{\vert \underbar{s}_{i}^{(m)}-s_{k(i)}^{(m)}\vert}-r\right)\Theta\left(\frac{\sigma}{r}-\vert \underbar{s}_{i}^{(m)}-s_{k(i)}^{(m)}\vert\right)}{\sum_{i=1}^{n-m-1}\Theta\left(\frac{\sigma}{r}-\vert \underbar{s}_{i}^{(m)}-s_{k(i)}^{(m)}\vert\right)},
\end{equation}
where $r$ is the threshold on the distance between two neighbouring points, $\Theta(\cdot)$ is the Heaviside step function
$$\Theta (x)=\begin{cases}0,&x<0\\1,&x\geqslant 0\, ,\end{cases}$$
$\underbar{s}_{k(i)}^{(m)}$ is the closest neighbour to $\underbar{s}_{i}$ in $m$ dimensions, $k(i)$ is the index $k\neq i$ of the time series element for which we have the minimum $\vert\underbar{s}_{i} - \underbar{s}_{k}\vert$,  and $\sigma$ is a parameter such that we remove from counting all the pairs of points whose initial distance is already large (precisely larger than $\sigma /r$). We make use of packages built for the R programming language: \emph{timeLag} from package ``nonlinearTseries'', which gives a criteria for estimating a proper time lag $\tau$; \emph{false.nearest} from package ``tseriesChaos'', which performs the method FNN to help deciding the smallest sufficient embedding dimension $m_d$.

\begin{figure}[tb]
\centering
\includegraphics[scale=0.60]{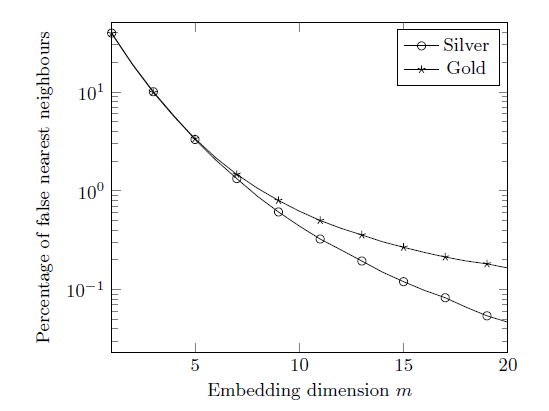}
\caption{Percentage of false nearest neighbours for metal futures series.}
\label{fig:fnn_metal_future}
\end{figure}

\begin{figure}[tb]
\centering
\includegraphics[scale=0.60]{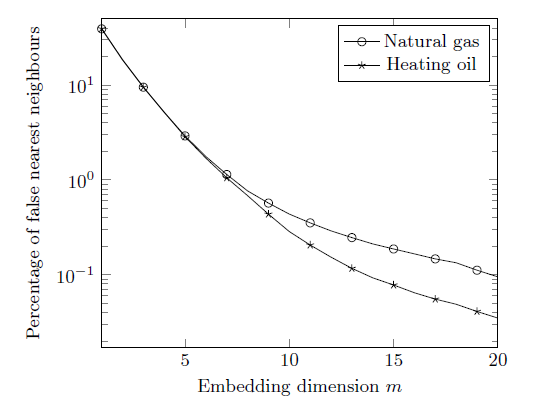}
\caption{Percentage of false nearest neighbours for energy futures series.}
\label{fig:fnn_energy_future}
\end{figure}

\begin{figure}[tb]
\centering
\includegraphics[scale=0.60]{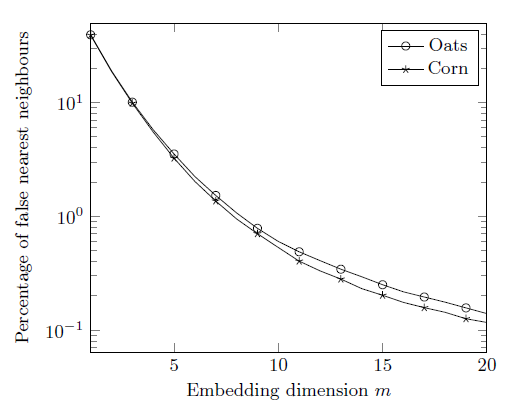}
\caption{Percentage of false nearest neighbours for grain futures series.}
\label{fig:fnn_grain_future}
\end{figure}

\begin{figure}[tb]
\centering
\includegraphics[scale=0.60]{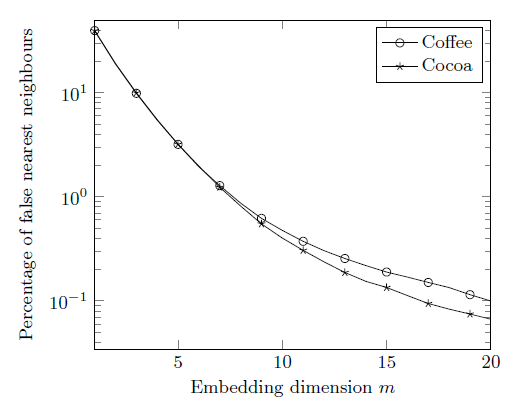}
\caption{Percentage of false nearest neighbours for soft commodity futures series.}
\label{fig:fnn_soft_future}
\end{figure}

\begin{figure}[tb]
\centering
\includegraphics[scale=0.60]{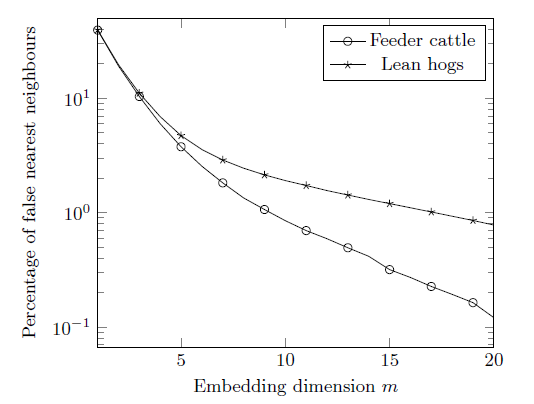}
\caption{Percentage of false nearest neighbours for agricultural  commodity futures series.}
\label{fig:fnn_agricoltural_future}
\end{figure}

In \figurename~\ref{fig:fnn_metal_future}, \figurename~\ref{fig:fnn_energy_future}, \figurename~\ref{fig:fnn_grain_future}, \figurename~\ref{fig:fnn_soft_future} and \figurename~\ref{fig:fnn_agricoltural_future} we plot the percentage of false nearest neighbours for the commodities considered in the paper, using the datasets described in Table \ref{table:dataset}. We adopt a logarithmic $y$-scale to make the differences between the curves visible.

%\begin{figure}
%\centering
%\figuretextsize
%  \begin{tikzpicture}
%  \begin{semilogyaxis}[black,
%  xlabel=Embedding dimension $m$,
%  ylabel=Percentage of false nearest neighbours,ymin=0,ymax=50,
%       xmin=1,
%       xmax=20]
%   \addplot[black, mark=o, mark repeat=2] table[col sep=comma] {FNN silver.csv};
%   \addlegendentry{Silver};
%   \addplot[black, mark=star, mark repeat=2] table[col sep=comma] {FNN gold.csv};
%      \addlegendentry{Gold};
%  \end{semilogyaxis}
% \end{tikzpicture}
%\caption{Percentage of false nearest neighbours for metal futures series.}
%\label{fig:fnn_metal_future}
%\end{figure}

%\begin{figure}
%\centering
%\figuretextsize
%  \begin{tikzpicture}
%  \begin{semilogyaxis}[black,
%  xlabel=Embedding dimension $m$,
%  ylabel=Percentage of false nearest neighbours,ymin=0,ymax=50,
%       xmin=1,
 %      xmax=20]
%   \addplot[black, mark=o, mark repeat=2] table[col sep=comma] {FNN natural.csv};
%   \addlegendentry{Natural gas};
%   \addplot[black, mark=star, mark repeat=2] table[col sep=comma] {FNN oil.csv};
%      \addlegendentry{Heating oil};
%  \end{semilogyaxis}
% \end{tikzpicture}
%\caption{Percentage of false nearest neighbours for energy futures series.}
%\label{fig:fnn_energy_future}
%\end{figure}

%\begin{figure}
%\centering
%\figuretextsize
%  \begin{tikzpicture}
 % \begin{semilogyaxis}[black,
%  xlabel=Embedding dimension $m$,
 % ylabel=Percentage of false nearest neighbours,ymin=0,ymax=50,
%       xmin=1,
 %      xmax=20]
  % \addplot[black, mark=o, mark repeat=2] table[col sep=comma] {FNN oats.csv};
%   \addlegendentry{Oats};
 %  \addplot[black, mark=star, mark repeat=2] table[col sep=comma] {FNN corn.csv};
  %    \addlegendentry{Corn};
%  \end{semilogyaxis}
% \end{tikzpicture}
%\caption{Percentage of false nearest neighbours for grain futures series.}
%\label{fig:fnn_grain_future}
%\end{figure}

%\begin{figure}
%\centering
%\figuretextsize
 % \begin{tikzpicture}
%  \begin{semilogyaxis}[black,
 % xlabel=Embedding dimension $m$,
%  ylabel=Percentage of false nearest neighbours,ymin=0,ymax=50,
 %      xmin=1,
  %     xmax=20]
%   \addplot[black, mark=o, mark repeat=2] table[col sep=comma] {FNN coffee.csv};
 %  \addlegendentry{Coffee};
  % \addplot[black, mark=star, mark repeat=2] table[col sep=comma] {FNN cocoa.csv};
%      \addlegendentry{Cocoa};
 % \end{semilogyaxis}
 %\end{tikzpicture}
%\caption{Percentage of false nearest neighbours for soft commodity futures series.}
%\label{fig:fnn_soft_future}
%\end{figure}

%\begin{figure}
%\centering
%\figuretextsize
 % \begin{tikzpicture}
%  \begin{semilogyaxis}[black,
 % xlabel=Embedding dimension $m$,
%  ylabel=Percentage of false nearest neighbours,ymin=0,ymax=50,
 %      xmin=1,
  %     xmax=20]
%   \addplot[black, mark=o, mark repeat=2] table[col sep=comma] {FNN feed.csv};
 %  \addlegendentry{Feeder cattle};
  % \addplot[black, mark=star, mark repeat=2] table[col sep=comma] {FNN lean.csv};
%      \addlegendentry{Lean hogs};
%  \end{semilogyaxis}
% \end{tikzpicture}
%\caption{Percentage of false nearest neighbours for agricultural  commodity futures series.}
%\label{fig:fnn_agricoltural_future}
%\end{figure}

If we set a threshold $\textrm{FNN}^{*}$, we use as embedding dimension the minimum value of $m$ such that $\textrm{FNN}<\textrm{FNN}^{*}$. For the threshold $\textrm{FNN}^{*}=0.5\%$, we get the embedding dimensions reported in Table \ref{table:minembed}.

%ATTENZIONE, PER IL LEAN HOGS SEI STATO COSTRETTO A SUPERARE M=20, COSA CHE NON SI EVINCE DALLA FIGURA \figurename~\ref{fig:fnn_agricoltural_future}

\begin{table}
\centering\begin{tabular}{lcc}
\toprule
Commodity contract & $\tau$ & $m$ \\
\midrule
Natural gas  &1 & 10\\
Heating oil &1 & 9\\
Gold &1 & 11\\
Silver &1 & 10\\
Corn &1 & 10\\
Oats &1 & 11\\
Cocoa &1 & 9\\
Coffee&1 & 10\\
Feeder cattle &1 & 13\\
Lean hogs &1 & 24\\
\bottomrule
\end{tabular}
\caption{Minimum embedding dimension (Threshold on FNN = 0.5\%)}
\label{table:minembed}
\end{table}

\subsection{Maximal Lyapunov exponent}

Measuring for sensitive dependence on initial conditions can be done by a mathematical operation using what are called Lyapunov Exponents. Let $\textbf{s}_{n_1}$ and $\textbf{s}_{n_2}$ be two points in state space with distance $\| \textbf{s}_{n_1}-\textbf{s}_{n_2}\|=\delta_0\ll 1$.

Denote by $δ_{\Delta n}$ the distance some time $\Delta n$ ahead between the two trajectories emerging from these points, $δ_{\Delta n}=\| \textbf{s}_{n_1+\Delta n}-\textbf{s}_{n_2+\Delta n}\|$. Then, the maximal Lyapunov exponent (MLE) $\lambda$ is determined by
\begin{equation}
\label{eq:lyap}
    δ_{\Delta n} \simeq \delta_0 e^{\lambda\Delta n}\, , \ \ \ \ δ_{\Delta n}\ll 1\, , \ \ \ \ \Delta n\gg 1 \, .
\end{equation}
If $\lambda$ is positive, this means an exponential divergence of nearby trajectories, i.e. butterfly effect. Naturally, two trajectories cannot separate farther than the size of the attractor, such that (\ref{eq:lyap}) is only valid during times $\Delta n$ for which $δ_{\Delta n}$ remains small. Otherwise, a saturation of the distance occurs and  therefore (\ref{eq:lyap}) is violated. Due to this fact, a mathematically more rigorous definition will have to involve a first limit $\delta_0 \rightarrow 0$ such that a second limit $\Delta n \rightarrow \infty$ can be performed without involving saturation effects. Only in the second limit does the exponent $\lambda$ become a well-defined and invariant quantity. For further details see \cite{kantz2004nonlinear}, Sec. 11.2.

\begin{figure}[tb]
\centering
\includegraphics[scale=0.60]{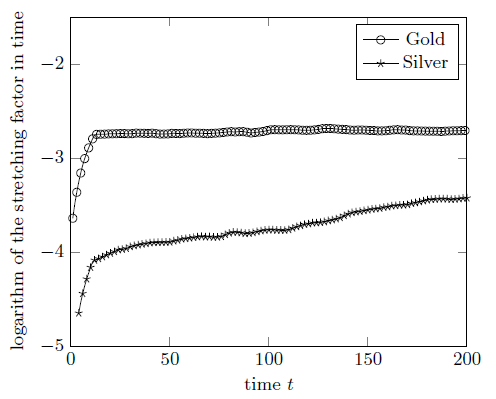}
\caption{Logarithm of the stretching factor in time for metal  commodity futures series.}
\label{fig:lyap_metal_future}
\end{figure}

\begin{figure}[tb]
\centering
\includegraphics[scale=0.60]{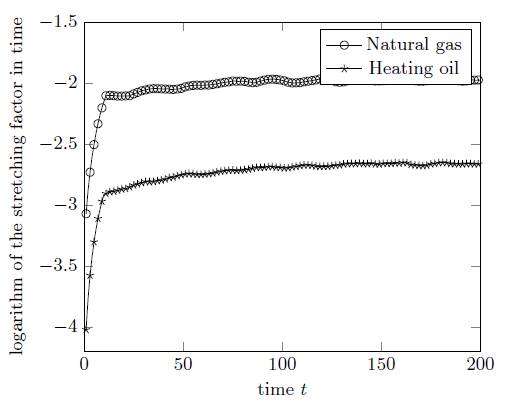}
\caption{Logarithm of the stretching factor in time for energy  commodity futures series.}
\label{fig:lyap_energy_future}
\end{figure}

\begin{figure}[tb]
\centering
\includegraphics[scale=0.60]{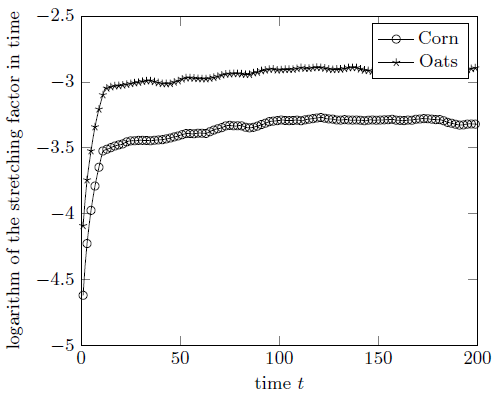}
\caption{Logarithm of the stretching factor in time for grain  commodity futures series.}
\label{fig:lyap_grain_future}
\end{figure}

\begin{figure}[tb]
\centering
\includegraphics[scale=0.60]{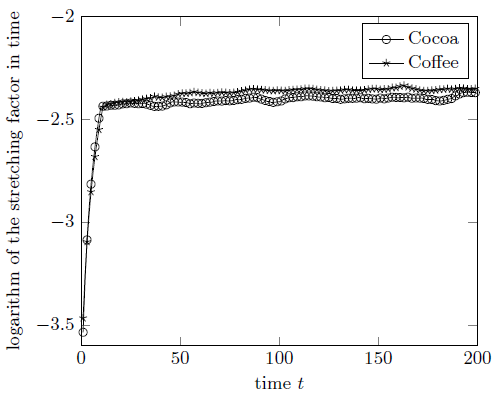}
\caption{Logarithm of the stretching factor in time for soft  commodity futures series.}
\label{fig:lyap_soft_future}
\end{figure}

\begin{figure}[tb]
\centering
\includegraphics[scale=0.60]{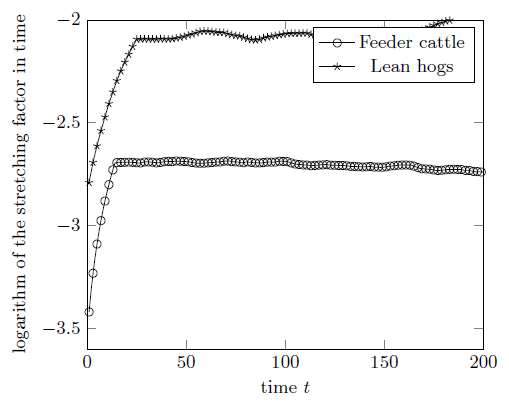}
\caption{Logarithm of the stretching factor in time for agricoltural  commodity futures series.}
\label{fig:lyap_agricoltural_future}
\end{figure}

To calculate $\lambda$, we used the routine R \emph{lyap\_k} from package ``tseriesChaos'', which performs the algorithm proposed in \cite{rosenstein1993practical} (see also \cite{hegger1999practical}).  \figurename~\ref{fig:lyap_metal_future}, \figurename~\ref{fig:lyap_energy_future}, \figurename~\ref{fig:lyap_grain_future}, \figurename~\ref{fig:lyap_soft_future} and \figurename~\ref{fig:lyap_agricoltural_future} depict the logarithm of the \emph{stretching factor} in time for the commodities considered in the paper. For the mathematical relationship between Lyapunov exponents and stretching factors see \cite{kantz2004nonlinear}, p. 204. If for some temporal range the logarithm of the stretching factor exhibits a robust linear increase, its slope is an estimate of $\lambda$ (see \cite{kantz2004nonlinear}, p. 70).

%% output <-lyap_k(series, m=, d=1, s=200, t=40, k=2, eps=4,ref=1700) %% varia m a seconda delle serie.

The empirical results obtained for maximal Lyapunov exponent $\lambda$ of each commodity are summarised in Table \ref{table:lambd}.

\begin{table}
\centering\begin{tabular}{lccc}
\toprule
Commodity contract & $\tau$ & $m$ & $\lambda$ \\
\midrule
Natural gas  &1 & 10& 0.08548951 \\
Heating oil &1 & 9& 0.09728674 \\
Gold &1 & 11& 0.07709301 \\
Silver &1 & 10& 0.09018312 \\
Corn &1 & 10& 0.09386063 \\
Oats &1 & 11& 0.08979586\\
Cocoa &1 & 9& 0.09346944\\
Coffee&1 & 10& 0.08772418\\
Feeder cattle &1 & 13& 0.05725445\\
Lean hogs &1 & 24& 0.03637373\\
\bottomrule
\end{tabular}
\caption{Maximal Lyapunov exponent $\lambda$ of each commodity.}
\label{table:lambd}
\end{table}

\subsection{Determinism test}

Since the commodity futures time series show an irregular random behaviour that often resembles chaos, we test the series in order to verify whether %the studied time series
it indeed originates from a deterministic system. For this purpose, we employ the determinism test introduced by \cite{Ka92}, using a package developed in \cite{Ko05}.

The test is based on the reconstruction of the state space from the observed variable. To construct an approximate vector field of the system, the phase space is covered by equally sized boxes with the same dimension $m$ as the reconstructed space. To each box, the average direction of the trajectory through the box during a particular pass is estimated. Each pass $i$ of the trajectory through the $k$-th box generates a unit vector, and the approximation for the vector field $\textbf{V}_k$ in the $k$-th box of the phase space is defined as the average vector of all passes. In \cite{Ka92}, $\kappa$ is defined as weighted average of $\textbf{V}_k$ with respect to the average displacement per step, $R_k^m$, of a random walk. In this way, the determinism coefficient $\kappa$ is equal to 1 for a deterministic system, while  $\kappa=0$ for a random walk. Then, for the intermediate cases, $\kappa$ measures the distance of time series from a deterministic system and from a stochastic process. The idea is to take the \emph{determinism coefficient} $\kappa$, obtained by Kaplan and Glass' test, as a measure of the reliability level (in percentage) of the test on sensitive dependence on initial conditions.
Actually, it is known  that time series generated from stochastic systems also may show positive MLE \cite{tanaka1996lyapunov}, \cite{ikeguchi1997lyapunov}, \cite{tanaka1998analysis}.

We use a package written in $C++$ code, which can be downloaded from M. Perc's Web page \footnote{M. Perc Web Page, \url{http://www.matjazperc.com/ejp/time.html}.}, as described in \cite{Ko05}.

\begin{table}
\centering\begin{tabular}{lcc}
\toprule
Commodity contract & $\tau$  & $\kappa$ \\
\midrule
Natural gas  &1 &   0.913584 \\
Heating oil &1 &   0.881473 \\
Gold &1 &   0.891292 \\
Silver &1 &  0.905958\\
Corn &1 &  0.904227 \\
Oats &1 & 0.915515\\
Cocoa &1 & 0.808978\\
Coffee&1 & 0.904908\\
Feeder cattle &1 & 0.904252\\
Lean hogs &1 & 0.932926 \\
\bottomrule
\end{tabular}
\caption{Determinism coefficient $\kappa$ of each commodity (2D projection).}
\label{table:lambda_kappa}
\end{table}

\begin{figure}
\centering
\subfloat[][\emph{Cocoa}]
{\includegraphics[scale=0.2]{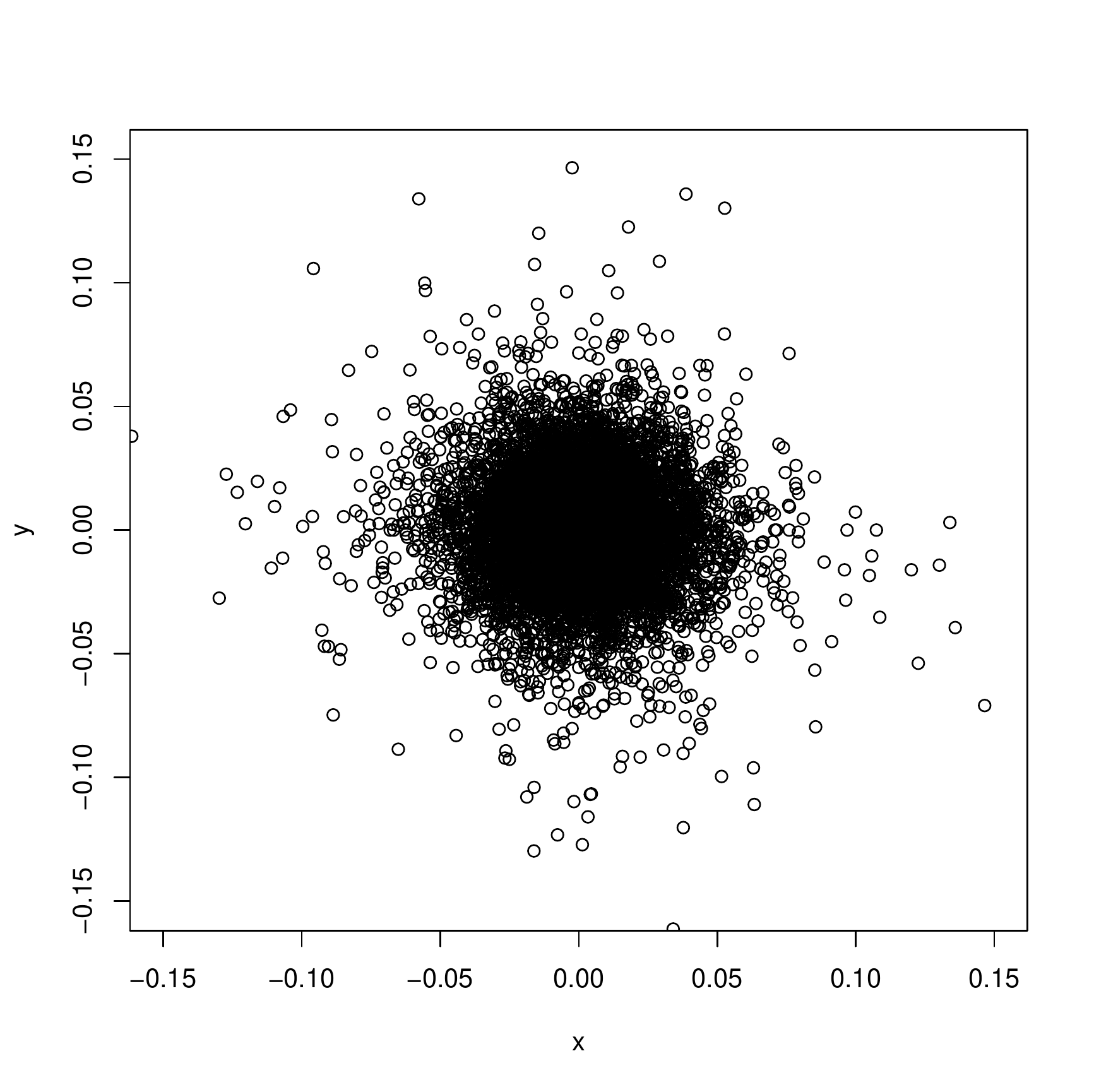}} \quad
\subfloat[][\emph{Coffee}]
{\includegraphics[scale=0.2]{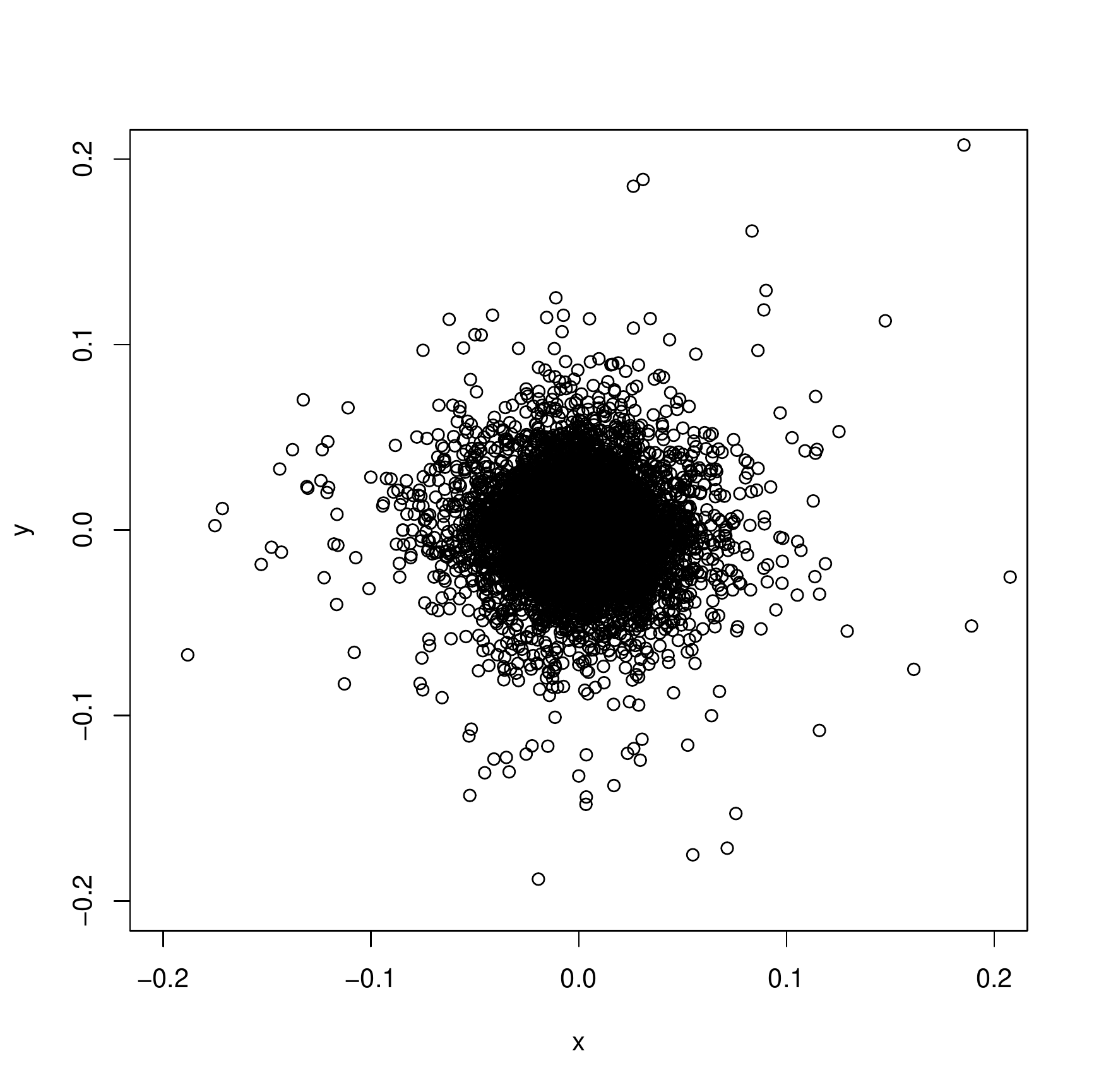}} \\
\subfloat[][\emph{Corn}]
{\includegraphics[scale=0.2]{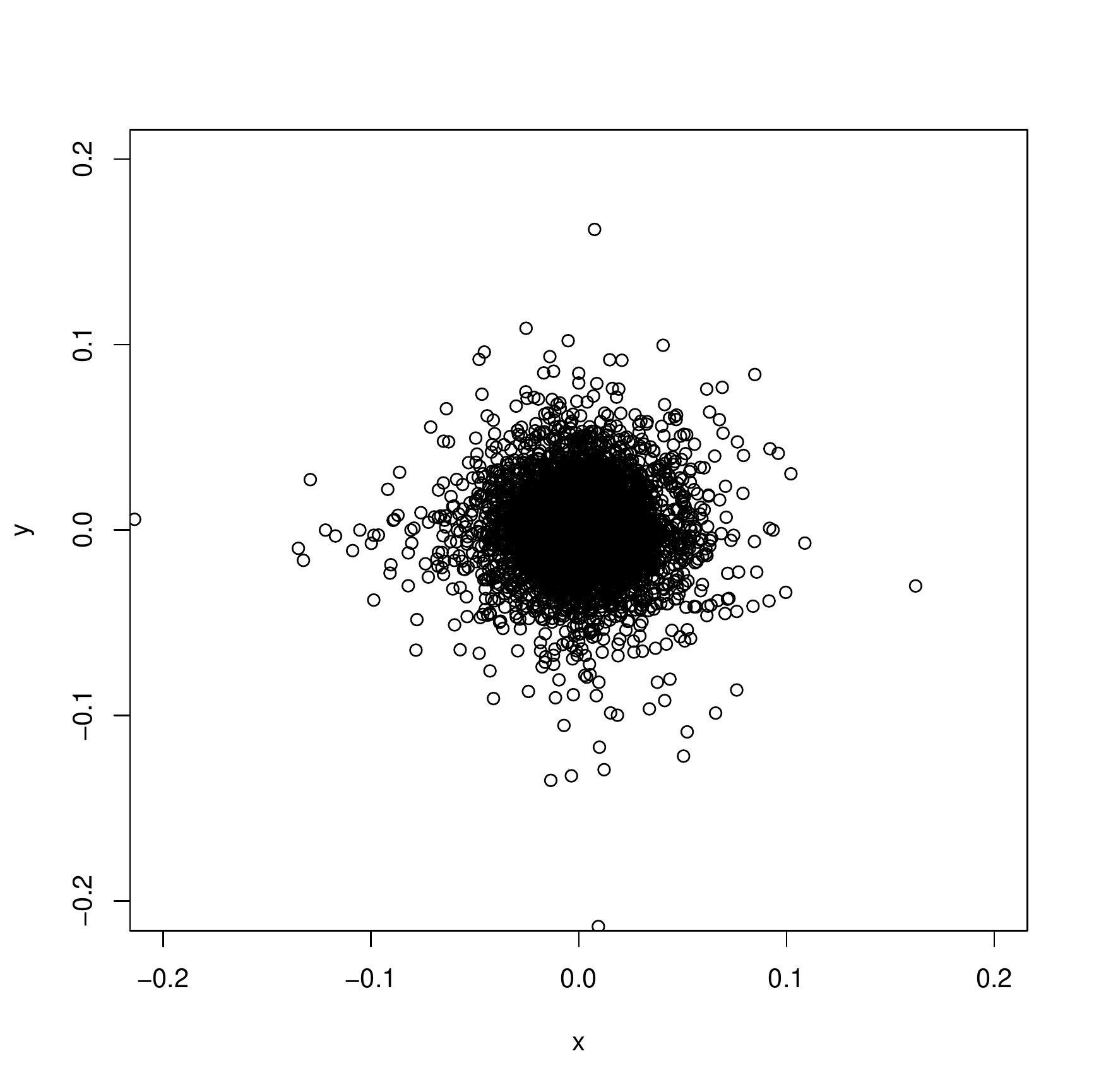}} \quad
\subfloat[][\emph{Feeder cattle}]
{\includegraphics[scale=0.2]{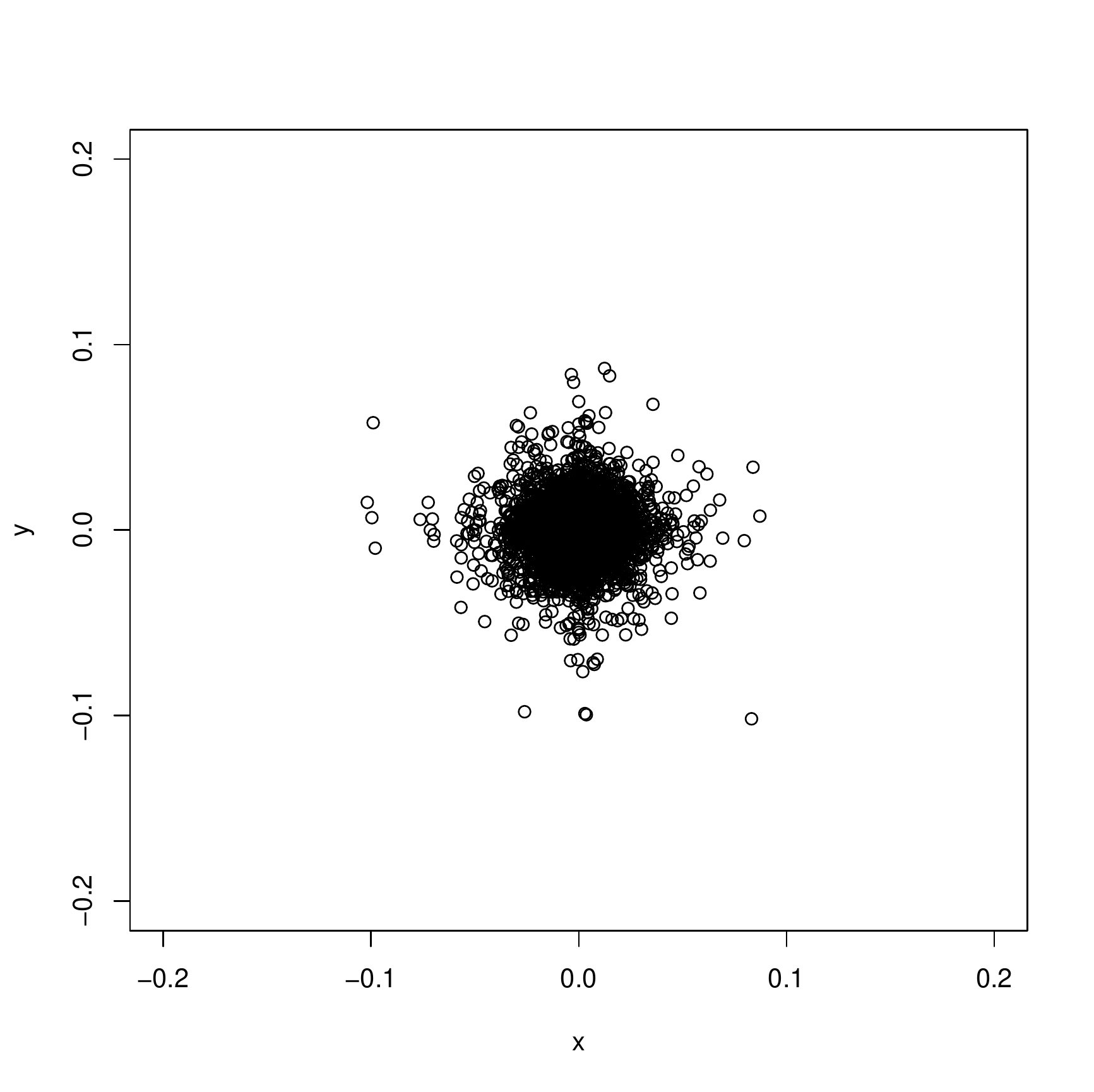}}
\caption{Reconstructed phase space of the cocoa, coffee, corn, feeder cattle future series (2D projection). Horizontal axis: $z_t$; vertical axis $z_{t+\tau}$.}
\label{fig:subfig1}
\end{figure}

\begin{figure}
\centering
\subfloat[][\emph{Gold}]
{\includegraphics[scale=0.2]{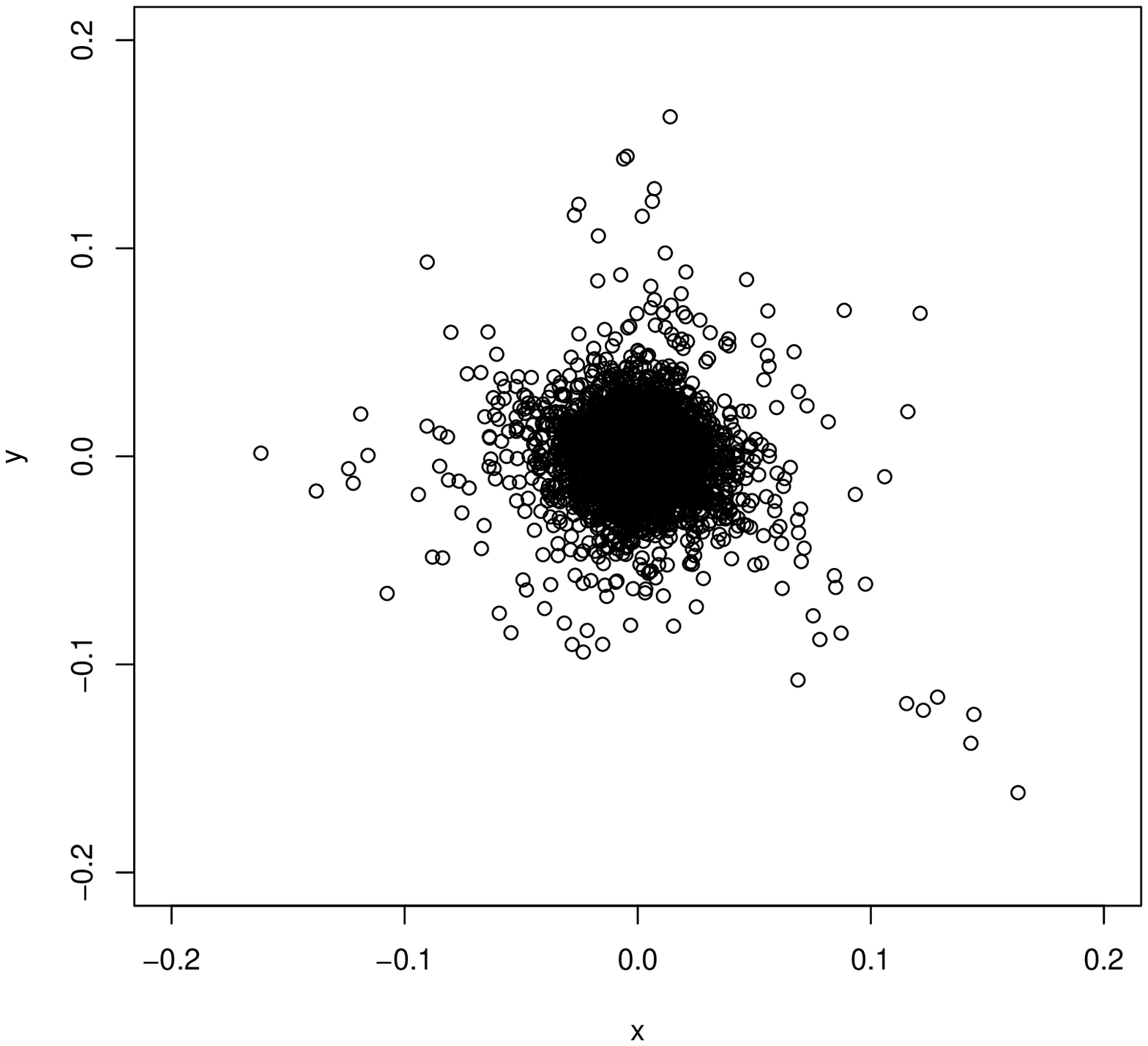}} \quad
\subfloat[][\emph{Lean hogs}]
{\includegraphics[scale=0.2]{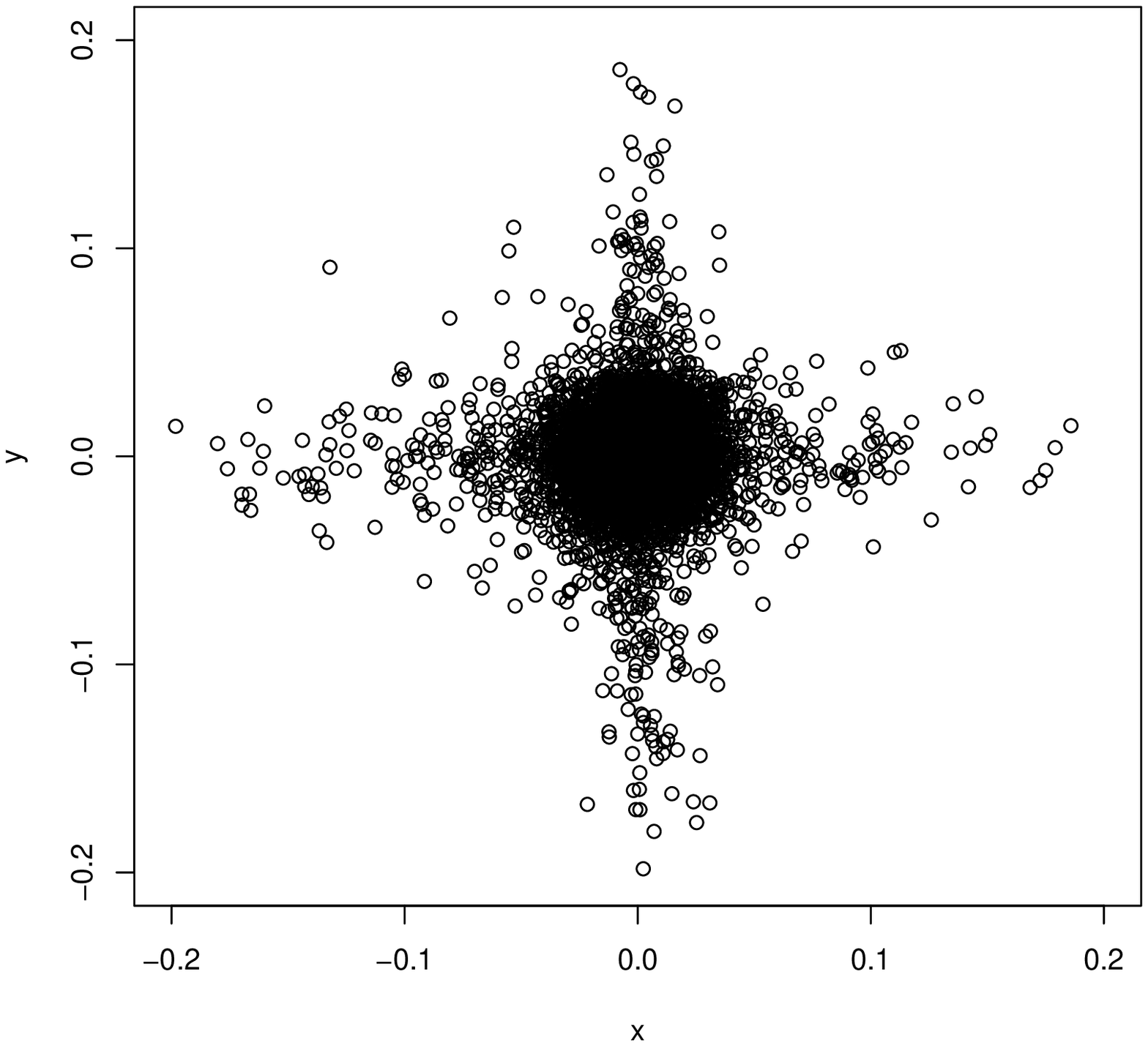}} \\
\subfloat[][\emph{Natural gas}]
{\includegraphics[scale=0.2]{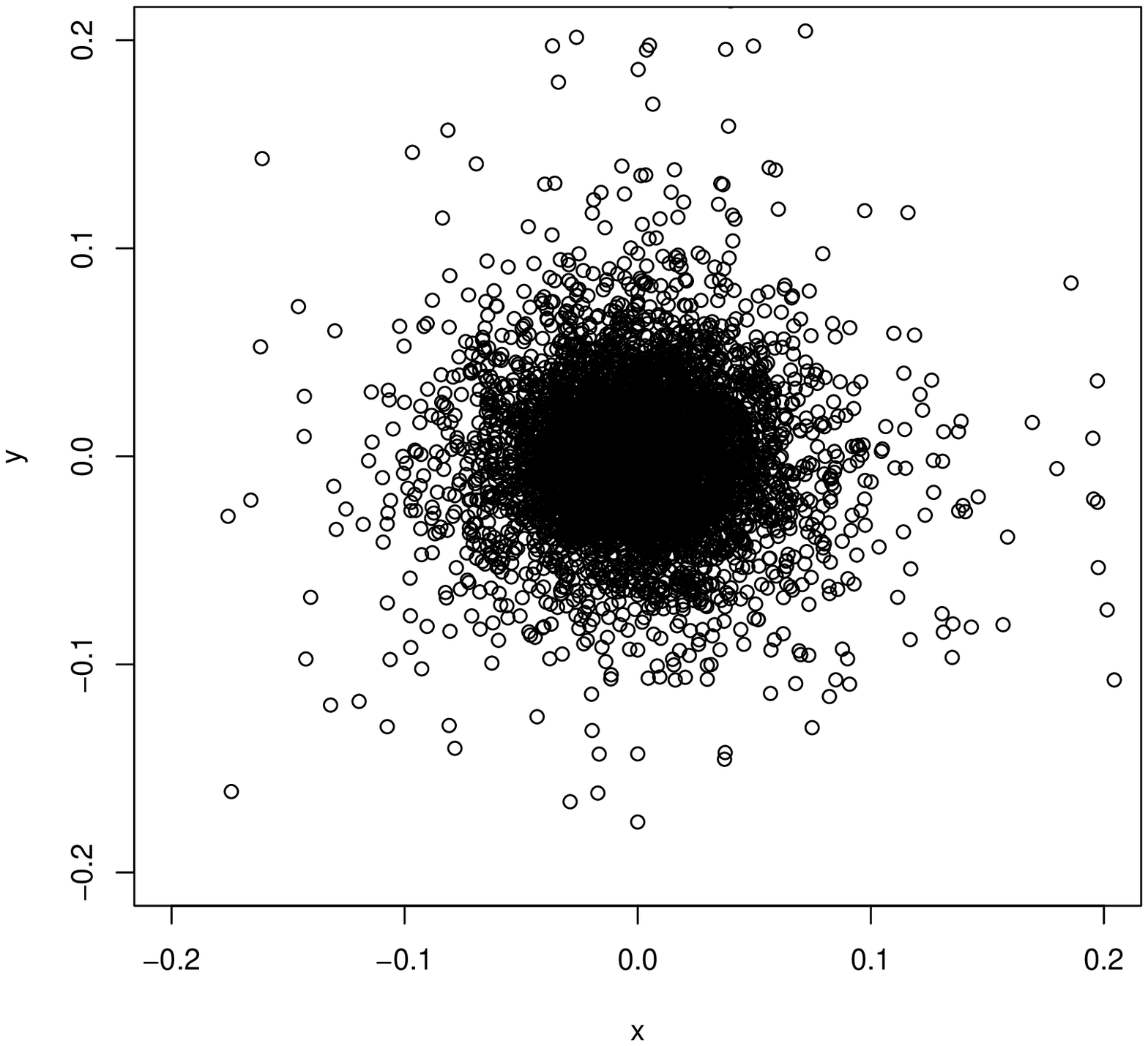}} \quad
\subfloat[][\emph{Oats}]
{\includegraphics[scale=0.2]{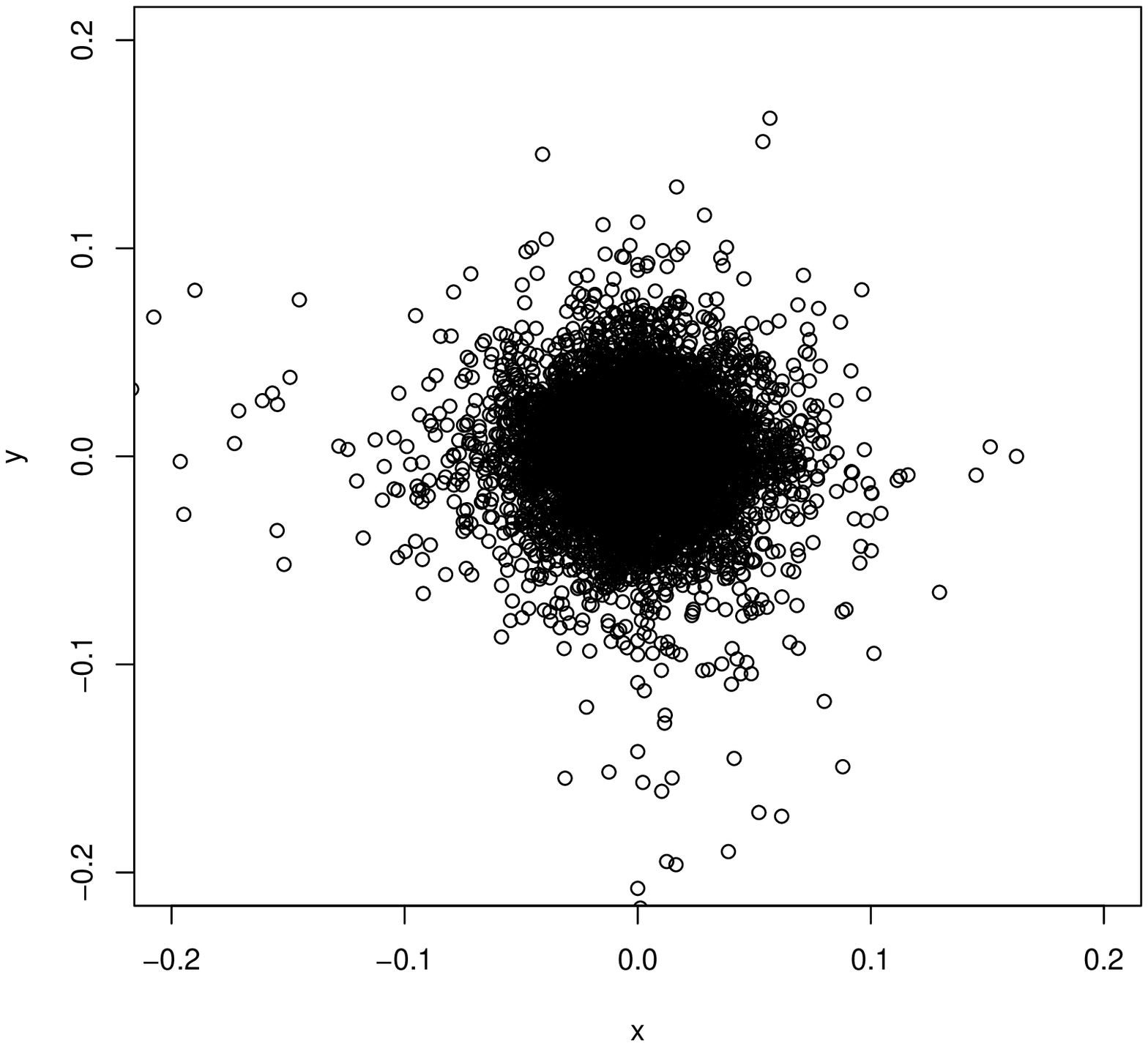}}
\subfloat[][\emph{Heating oil}]
{\includegraphics[scale=0.2]{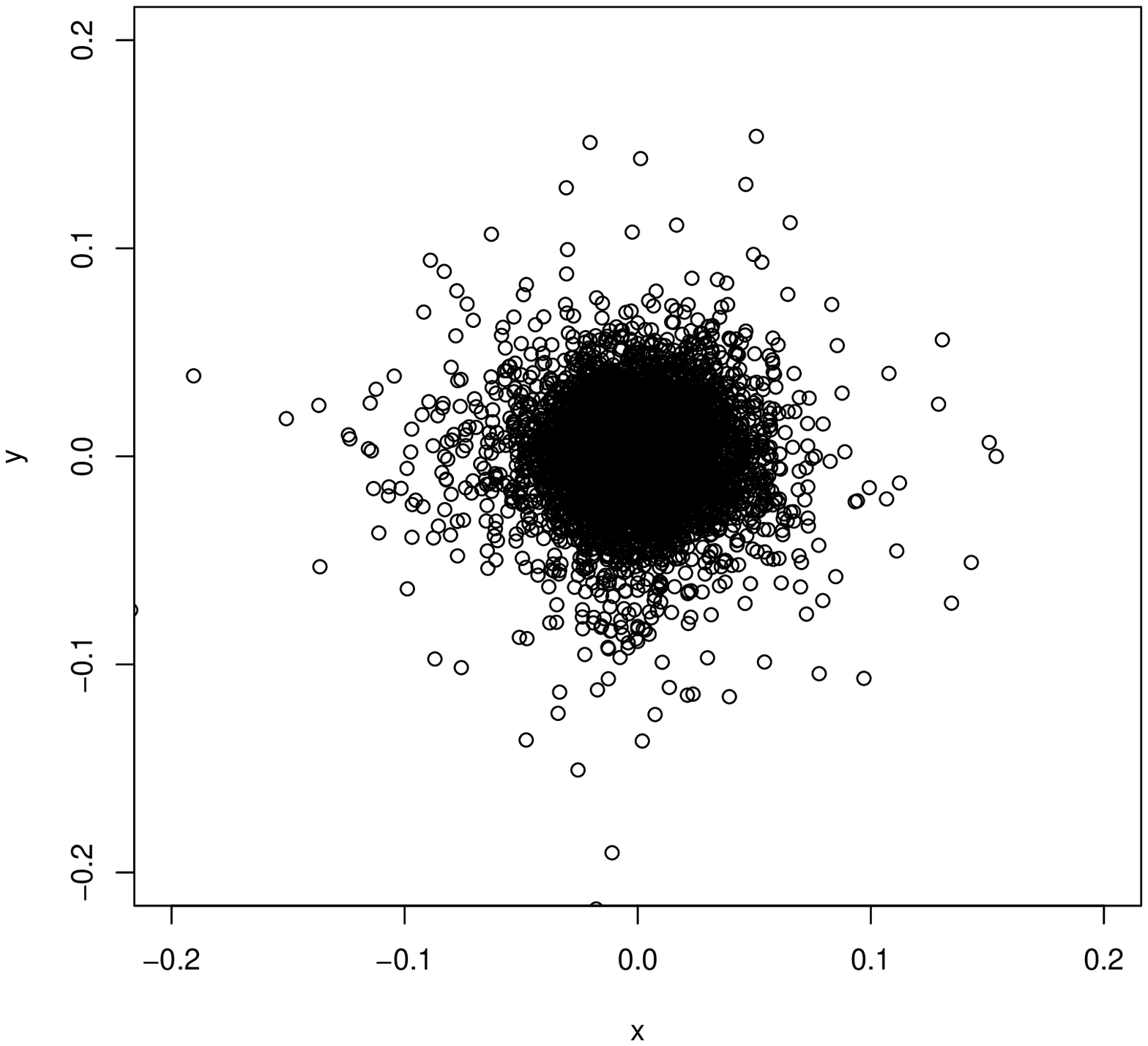}}
\subfloat[][\emph{Silver}]
{\includegraphics[scale=0.2]{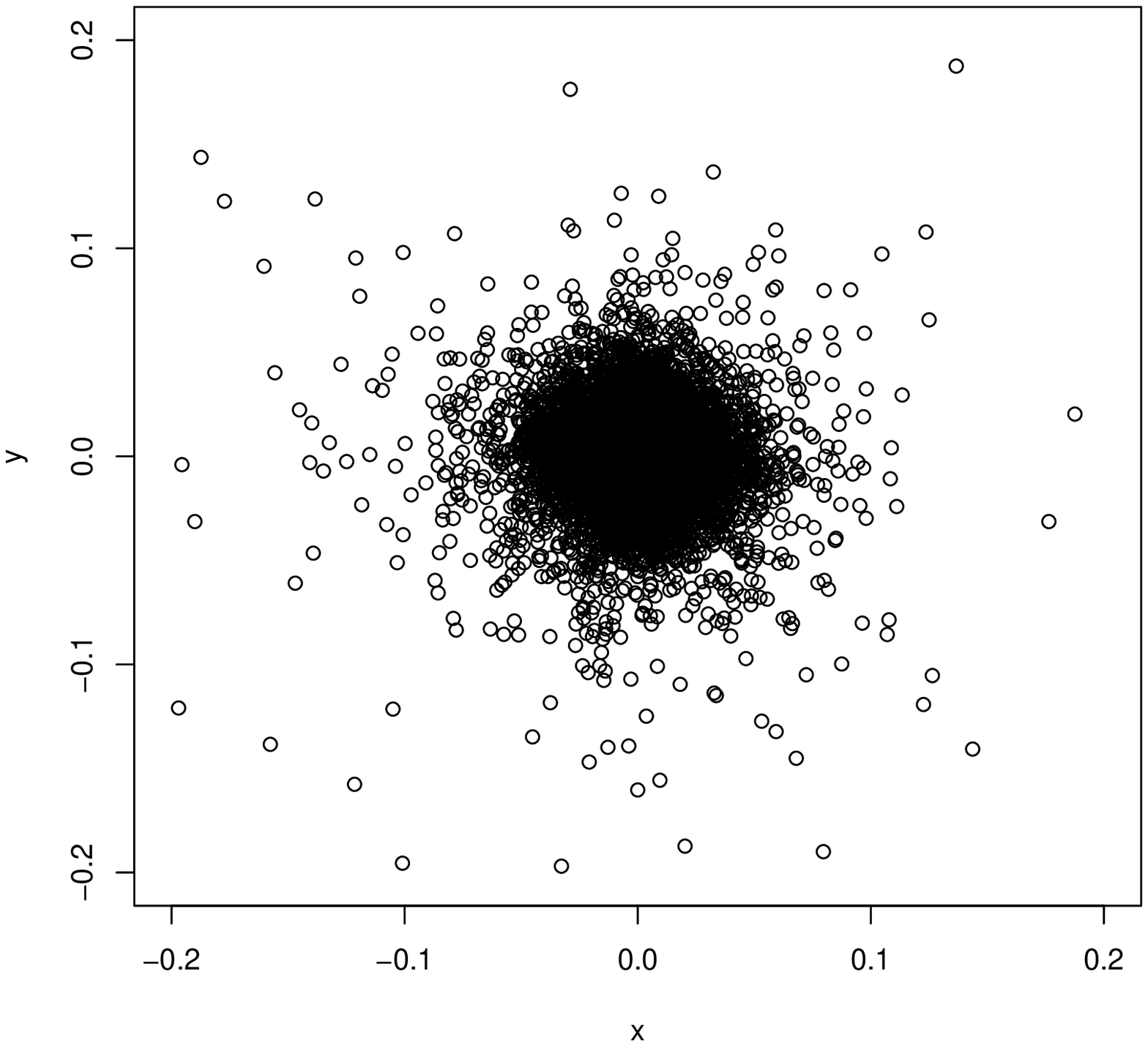}}
\caption{Reconstructed phase space of the gold, lean hogs, natural gas, oats, heating oil and silver future series (2D projection). Horizontal axis: $z_t$; vertical axis $z_{t+\tau}$.}
\label{fig:subfig2}
\end{figure}

The results of the determinism test for the embedding space presented in \figurename~\ref{fig:subfig1} and \figurename~\ref{fig:subfig2} are shown in Table \ref{table:lambda_kappa}. For this calculation the multidimensional embedding space was coarse grained into a $\underbrace{25 \times 25 \times\dots \times 25}_m$ grid. In Table \ref{table:lambda_kappa}, $\kappa$ is near to $1$ for all  commodity futures series. So, the deterministic signature is still good enough to be preserved, so that the chaotic appearance of the reconstructed attractor cannot be attributed to stochastic influences.

See \figurename~\ref{fig:subfig1} and \figurename~\ref{fig:subfig2}. When we look only at the data, we think it were coming out of a black box. If the coming was regular, the graph would be suitably dull: every point would land at the same place. The graph would be a single dot (or almost). But the commodity futures series are, in general, subject to noise and here sensitive to initial conditions (Table \ref{table:lambd}). As we sayed, it is not a predictable pattern beyond a short time. So instead of the single dot, we will see in \figurename~\ref{fig:subfig1} and \figurename~\ref{fig:subfig2} a slightly fuzzy blob. If it were truly random, points would be scattered all over the graph. There would be no relation to be found between one interval and the next. But if a strange attractor were hidden in the data, it might reveal itself as a coalescence of fuzziness into distinguishable structures.

%Citare: The Dripping Faucet as a Model Chaotic System, Shaw

%%%%%%%%%%%%%%%%%%%%%%%%%%%%%%%%%%%%%%

\section{A comparison with related results}
\label{sec:disc}

In this paper we have adopted an approach which can be schematized in the following points: 1) reconstruction of the phase space, where we estimated the smallest sufficient embedding dimension using the FNN algorithm. Moreover, on the basis of the parameters of previous point: 2) we have tested for sensitive dependence on initial conditions using Lyapunov Exponents; 3)  we have estimated the determinism coefficient of each time series.

Between the references considered here, the estimation of smallest sufficient embedding dimension $m$ is performed by methods for determining it only in \cite{matilla2007nonlinear}, \cite{sakai2007transition}, \cite{barkoulas2012metric}. Other references considered $m$ arbitrarily varying on a certain range (for example, $m\in[1,6]$ in \cite{kyrtsou2009energy} and $m\in[3,10]$ in \cite{chwee1998chaos}), but a similar approach does not allow to select an appropriate value for $m$. In fact, as explained in \cite{Pack80, Take81, Kennel92, ruelle1989chaotic}, the FNN procedure identifies the number of ``false nearest neighbors'', points that appear to be nearest neighbours because the embedding space is too small, of every  point on the attractor associated with the set $\{\textbf{p}(i),\, i= 1,2,\dots, n-(m-1)\tau\}$, as defined in (\ref{eq:3}). When the number of false nearest neighbors drops to zero, we have unfolded or embedded the attractor in $m$-dimensional Euclidian space. If we chose $\bar m< m$, i.e. a dimension less than the appropriate embedding dimension $m$, we are viewing the attractor in too small an embedding space and an estimation of the MLE for an embedding dimension $\bar m$ might yield misleading results about the sensitive dependence on initial conditions of the analyzed time series. The same conclusion might occurs in the case $\bar m> m$. In fact, from a mathematical point of view of the embedding process it does not matter whether one uses the minimum embedding dimension $m$ or any $\bar m> m$, since once the attractor is unfolded, the theory's work (see \cite{Take81}) is done. But, in more relatable terms, working in any dimension larger than the minimum required by the data leads to the problem of contamination by roundoff or instrumental error since this ``noise'' will affect the additional $\bar m- m$ dimensions of the embedding space where no dynamics is operating \cite{Kennel92}.

%First of all, we emphasize that the maximum Lyapunov exponent test offers a clearer result than Kolmogorov entropy. In fact, if the maximum Lyapunov exponent is positive, butterfly effect would be implied; on the contrary, it would not. On the other side, $K >0$ is only a sufficient condition for butterfly effect.

Even if the results obtained with the MLE indicate the presence of butterfly effect (Table \ref{table:lambd}), it is anyway important to compare the reliability of the results obtained about sensitive dependence on initial conditions.
As we already said, some stochastic systems may show sensitive dependence on initial conditions (\cite{tanaka1996lyapunov}, \cite{ikeguchi1997lyapunov}, \cite{tanaka1998analysis}); thus we propose to assume the results concerning the determinism coefficient $\kappa$ as the reliability level (in percentage) of the MLE.

Some papers (\cite{lai2016dynamic}, \cite{matilla2007nonlinear}, \cite{Panas01}, \cite{Panas00}, \cite{sakai2007transition}, \cite{serletis1999north}) claim to have discovered the evidence of butterfly effect. \cite{lai2016dynamic} uses the BDS test and estimates the correlation dimension employing Grassberger-Procaccia method. \cite{Panas00} and \cite{Panas01} employ the BDS test and estimate the Kolmogorov entropy, the correlation dimension and the MLE. \cite{serletis1999north} work in order to obtain stationary and appropriately filtered data, removing any linear as well as nonlinear stochastic dependence, and then estimate the MLE. Although these studies investigate the presence of both stochastic and determinist components in time series, they do not provide any estimate of the determinism rate existing in the analyzed data. On the other hand, among the cited references, \cite{matilla2007nonlinear} and \cite{sakai2007transition} come closest %sono quelle che si avvicinano maggiormente
to the approach followed in this paper. They both build upon a test on determinism of the analyzed time series.

Matilla-Garc\'{i}a employs a test developed by \cite{KAPLAN199438}, which introduces a coefficient $K$. The nonzero value of $K$ is interpreted as the level of nondeterminism in the data. However, a range $K\in(0,\infty)$ is not very specific. For instance, it does not detects the level of determinism in the analyzed time series, it only assesses its order when compared with others.

The work of \cite{sakai2007transition} is the only one that uses a deterministic test as we do here. %, despite authors focus on a future series which we have not considered (the piglet-pricing data).
Their deterministic investigation is based on the test developed by \cite{wayland1993recognizing}. In this test, the quantity that provides information about the determinism of the time series is the \emph{median translation error}. If this error is close to 0, then the time series is the result of a deterministic process; if the median translation error is close to 1, then the time series is the result of a stochastic process. In \cite{sakai2007transition}, the translation error is very close to $0$, which suggests a very high level of determinism in the time series.

%With regards to the futures time series analyzed in the present paper, we make the following comparison with respect to other works.

The presence of butterfly effect in commodity futures markets is a controversial matter, as can be deduced from the literature reviewed in Section \ref{sec:review}: %for similar markets,
some papers claim they have detected the presence of chaos (butterfly effect), while others state the opposite. Let us now compare our results with those of others, with regards to the futures time series we have presently analyzed.

\cite{kyrtsou2009energy} have evaluated natural gas and heating oil, over the period from 1994 to mid-January 2008, showing that Lyapunov exponent estimates are negative. They also show the existence of a structure that is partially deterministic. In our paper, even if we have considered different commodities or different kind of series (futures or spot prices), the futures time series show a considerable contribute of determinism ($\kappa$ near to $1$ for all the commodities), which differs from the observation of the partial determinism of the structure enlightened by \cite{kyrtsou2009energy}. Moreover, in the case of natural gas and heating oil, the MLE that we have detected are positive, while those estimated by \cite{kyrtsou2009energy} are negative: these two approaches give different results on Lyapunov exponent. However, we point out that: 1) the approach followed by \cite{kyrtsou2009energy} is different from that employed here and, in fact, they test for chaos by applying methods based on neural networks; 2) the time series do not match because they use daily spot prices, provided by \url{www.barchart.com}; 3) the sample periods are different, because \cite{kyrtsou2009energy} consider the period by 3.1.1994 to 25.1.2008 while we have examined the ranges 06.03.1979 - 15.05.2014 for heating oil and 03.04.1990 - 15.05.2014 for natural gas. A comparison of these two approaches and a new set of measures conducted on the same temporal range could be a useful exercise for next work.

As for heating oil, \cite{Adrangi01}, for observations on the range 1/02/85 - 03/31/95, employ correlation dimension test, the BDS test and Kolmogrov entropy, without finding evidence of butterfly effect.

\cite{matilla2007nonlinear} uses observations of natural gas futures princes, starting from 04/03/1990 to 10/19/2005. He discovers the positivity of the MLE, wheras for the same energy commodity \cite{chwee1998chaos}, examining observations from April 1990 to September 1996, shows no evidence of butterfly effect from the estimation of the Lyapunov spectra.
The results on the positivity of MLE obtained by \cite{matilla2007nonlinear} are in accordance with ours. This is not surprising, since Matilla-Garc\'{i}a employs the same method (by \cite{rosenstein1993practical}) used here.

As for natural gas \cite{serletis1999north} examining monthly data from 1985:2 to 1996:12, ``test for positivity of the dominant Lyapunov exponent. Before conducting such a nonlinear analysis, the data were rendered stationary and appropriately ﬁltered, in order to remove any linear as well as nonlinear stochastic dependence''. They find evidence of butterfly effect in all natural gas liquids markets. Differently from their paper, we do not apply any filtering to data.

As for corn, Chatrath, et al. \cite{Chatrath02} examine data from 11 December 1969 to 30 March 1995. They employ three tests: the correlation dimension, the BDS statistic and a measure of Kolmogorov entropy. These methods reveal that there is no consistent evidence of low dimension chaos in commodity futures prices.

We recall that all the papers cited (\cite{adrangi2002dynamics}, \cite{Adrangi01}, \cite{barkoulas2012metric}, \cite{Chatrath02}, \cite{chwee1998chaos}, \cite{kyrtsou2009energy}, \cite{lai2016dynamic}, \cite{matilla2007nonlinear}, \cite{moshiri2006forecasting}, \cite{Panas01}, \cite{Panas00}, \cite{sakai2007transition}, \cite{serletis1999north}) investigate the ``experimental'' definition of chaos. Among them, \cite{lai2016dynamic}, \cite{matilla2007nonlinear}, \cite{Panas01}, \cite{Panas00}, \cite{sakai2007transition}, \cite{serletis1999north} claim to have discovered the evidence of ``chaos'' but this term is abused: all the authors cited above investigate the butterfly effect, that is only one of the properties of a chaotic system (see Definition of chaos in \cite{Devaney89}). %In any case, establish whether a data series shows a deterministic butterfly effect represents a not easy task, because some time series generated from stochastic systems might show sensitive dependence on initial conditions.
It is not a negligible particular. As they rightly say, ``the theory is practice'', to mean that the effectiveness of a theory is based on its ability to generate a knowledge of phenomena so accurate as to allow the formulation of reliable forecasts. This also means that an insufficient regard to theoretical aspects might yield results that are not reliable. The contrasting results about the presence of butterfly effect in commodity futures markets may be generated, for example, by this misunderstanding.

For a deeper discussion of the mathematical aspects of the chaos definition we talked about here, see  \cite{MasVel2}.

\section{Conclusions}
\label{sec:conc}

We have reviewed and analysed the presence of butterfly effect for several commodities markets. In particular, we focused on the following commodity futures series: two energy series (natural gas, heating oil), two metal series (gold, silver), two grains (corn, oats), two soft commodities (cocoa, coffee) and two other agricultural commodities (feeder cattle, lean hogs) futures from the Chicago Board of Trade (CBOT), Chicago Mercantile Exchange (CME), Inter Continental Exchange (ICE), New York Mercantile Exchange (NYMEX), and its division Commodity Exchange (COMEX). The empirical results obtained in the above analysis are summarised in Tables \ref{table:minembed}, \ref{table:lambd}, \ref{table:lambda_kappa}.

%We have employed a package developed in \cite{Ko05} and based on the determinism test introduced in \cite{Ka92}.
%For each commodity we have analyzed the determinism rate $\kappa$ on varying of $\tau$. Afterward, in order to check sensitive dependence on initial conditions we have selected the largest Lyapunov exponents tests, using a Matlab code based on algorithms presented, respectively, in \cite{pincus1991approximate} and \cite{Wolf85}. The empirical results obtained in the above steps are summarised in  ....

We used the Lyapunov Exponents and a determinism test, both based on the reconstruction of the phase space. In particular, we employed a coefficient $\kappa$ that describes the \emph{determinism rate} of the analyzed time series. The coefficient $\kappa$ represents, in percentage, the reliability level about the test on the sensitive dependence on initial conditions. The introduction of this reliability level is motivated by the fact that time series generated from stochastic systems might show sensitive dependence on initial conditions. The coefficient $\kappa$ has been introduced by \cite{Ka92}, and we employ here a recent methodology developed by \cite{Ko05}.

In our work, the reliability level $\kappa$ yields results near to $1$ while the MLE is positive for all  commodity futures series. This means that the latter show a considerable contribution of determinism. In this way, we can ensure the presence of butterfly effect (that is one of the properties of a chaotic system, according to \cite{Devaney89}) in the commodity futures markets considered in the paper.

The results of the empirical treatments that we present here are, of course, subject to several interpretations.

The magnitudes of the Lyapunov exponents quantify a system's dynamics in information theoretical terms. The exponents represent the rate at which the system creates or distorts information (\cite{shaw1981strange}). For this purpose, let us consider a unit measuring error in all the  commodity futures series, and consider, as an instance, oats and corn commodities since they share the same temporal range of observation. Denote, respectively, with $\lambda_{o}$ and $\lambda_{c}$ their MLEs, then the error in the corn series has been amplified $\lambda_c/\lambda_o\simeq0.09386/0.08979\simeq 1.0452$ times faster than in the oats series.

The role of information in the markets, as we know, is of paramount importance. Let us think of the formalization proposed by \cite{malkiel1970efficient}, \cite{Fama1998283}. The central issue is whether or not to adopt trading strategies that achieve excess returns relative to the market, based on information contained in the historical data. Currently, the empirical evidence would seem to indicate that markets are often not efficient, even in weak form. The perception of a trend as seemingly stochastic could be due to the lack of knowledge of the information underlying it.

In other words, for the commodities considered in this paper the empirical analysis suggests that there are several deterministic forces interacting with each other. The presence of a chaotic dynamics could be connected to the existence of several deterministic forces that may result in complex price movements in financial markets. Today, the complexity of financial and commodities markets is very high because world decisions in business, finance and economics are influenced by sociologic, environmental, and geopolitical factors. In this regard, Panas and Ninni \cite{Panas00} write in their conclusions: ``An energy economist who is interested in the dynamic behaviour of the complex system that governs the oil markets needs to know how sensitive the system is to initial conditions, and to achieve this he needs to estimate the Lyapunov exponents''.

Thanks to recent methodologies (e.g. package developed by \cite{Ko05}),
we prove that this is agreeable also in other commodity and energy markets, but it is useful to spell out the conditions under which that is possible. We can add that an economist interested in complex system dynamic behaviour needs to know how deterministic the system is as well as how sensitive it is to initial conditions. That is why he needs to estimate the determinism coefficient and the MLE.

\section*{Acknowledgements}

The authors thank Prof. Matja\u{z} Perc for the C++ code of the package developed in \cite{Ko05}.

\section*{References}

\bibliographystyle{te}
\bibliography{energy}

\end{document}